

\PHYSREV
\overfullrule0pt
\def\lr{SU(2)_L\times SU(2)_R}
\def\hl{h^0}
\def\hh{H^0}
\def\ha{A^0}
\def\cosb{\cos\beta}
\def\sinb{\sin\beta}

\def\ie{{\it i.e.}}

\def\SCIPP{\centerline {\it Santa Cruz Institute for Particle Physics}
\vskip -.4pt
  \centerline{\it University of California, Santa Cruz, CA 95064}}
\def\SLAC{\centerline {\it Stanford Linear Accelerator Center}
\vskip -.4pt
          \centerline {\it  Stanford University, California  94309} }
\def\SMU{ \centerline {\it Department of Physics,
          Southern Methodist University }
\vskip -.4pt
          \centerline {\it Dallas, Texas 75275-0175} }
\def\ifmath#1{\relax\ifmmode #1\else $#1$\fi}
\def\bold#1{\setbox0=\hbox{$#1$}%
     \kern-.025em\copy0\kern-\wd0
     \kern.05em\copy0\kern-\wd0
     \kern-.025em\raise.0433em\box0 }

 \doublespace
{\singlespace
 \Pubnum{SLAC--PUB--6196\cr
SCIPP--93/08\cr
SMU HEP 93--06}
 \pubtype{T/E}
\date{May 1993}
\titlepage

\title{{\fourteenbf Constraints on CP violation in the Higgs sector
from the $\rho$ parameter}\doeack}
\author{Alex Pomarol}
\SCIPP
\andauthor{Roberto Vega}
\SLAC
\centerline{and}
\SMU

\vfill

{\singlespace
\centerline{\bf Abstract}
We discuss the relation between the
CP symmetry and the custodial $SU(2)$ symmetry
in the Higgs sector. In particular,
we show that CP violation in the Higgs-gauge  sector is allowed
only if
 the custodial
$SU(2)$ symmetry is broken.  We exploit these facts  to constrain CP
 violation using the experimental bounds on $\rho-1$.
CP nonconservation in the Higgs-fermion interactions can also be
constrained in a similar way although a possible  exception is
pointed out.  \par
}

\vfill
\submit{Physical Review D.}
\vfill
\endpage

\noindent{\bf 1. Introduction}

\REF\kob{M. Kobayashi and M. Maskawa, Prog. Theor. Phys.
 {\bf 49} (1973) 652.}

In the standard model  (SM) of the weak interactions the
origin of CP nonconservation is  intimately related to the mechanism
of electroweak symmetry  breaking (ESB).
Progress in the understanding of one of these phenomena  will most
probably lead to progress in the  understanding of the other.
\REF\wei{S. Weinberg, Phys. Rev. {\bf D42} (1990) 860.}
\REF\weib{S. Weinberg, Phys. Rev. Lett. {\bf 63} (1989) 2333.}
\REF\gun{J.F. Gunion, D. Wyler, Phys. Lett. {\bf B248} (1990) 170.}
\REF\bar{S.M. Barr and A. Zee, Phys. Rev. Lett. {\bf 65} (1990) 61;
J.F. Gunion and R. Vega, Phys. Lett. {\bf B251} (1990) 157;
R.G. Leigh et al., Nucl. Phys. {\bf B352} (1991) 45;
D. Chang, W.-Y. Keung and T.C. Yuan, Phys. Rev. {\bf D43} (1991) 14;}
\REF\grz{B. Grzadkowski and J.F. Gunion, Phys. Lett. {\bf B287} (1992)
237;
C. Schmidt and M. Peskin, Phys. Rev. Lett. {\bf 69} (1992) 410;
R. Cruz, B. Grzadkowski and J.F. Gunion, Phys. Lett. {\bf B289}
 (1992) 440;
D. Atwood, G. Eilam and A. Soni, Phys. Rev. Lett. {\bf 70} (1993) 1364
; preprint SLAC-PUB-6083.}
\REF\men{A. M\'endez and A. Pomarol, Phys. Lett. {\bf B272} (1991) 313.}
\REF\grzb{B. Grzadkowski and J.F. Gunion, Phys. Lett. {\bf B294} (1992)
361.}
\REF\pil{A. Pilaftsis and M. Nowakowski, preprint MZ-TH-92-56 (1992);
A. Soni and R.M. Xu, preprint BNL-48160 (1992);
D. Chang, W.-Y. Keung and I. Phillips, preprint CERN-TH-6814-93 (1993);
X.-G. He, J.P. Ma and B. McKellar, preprint UM-P-93-11 (1993).}

CP nonconservation in the SM arises from
the phases appearing in the Kobayashi-Maskawa
matrix\refmark{\kob}.  This suffices to
explain  the CP violation observed
in the neutral kaon system.
However,  unless only one Higgs doublet triggers  ESB, it seems
natural  for CP nonconservation  to also occur in the Higgs
interactions\refmark\wei.
This source of CP violation can manifest itself
through virtual effects  or through direct production of  Higgs bosons.
In the first case, one looks for one-loop Higgs contributions
to CP-violating experimental observables such as the
 electric dipole moment of the neutron\refmark{\wei-\gun}\
  and electron\refmark{\bar}, or
asymmetries in the top quark production or decay\refmark\grz.
  Such  one-loop contributions are generally small
and decrease with increasing Higgs masses. For Higgs masses of
the order of a hundred GeV maximal CP violation in the Higgs sector
is  consistent with present experimental data.
In the second case, one tries to detect CP violation in the production
mechanisms of the Higgs bosons or their decays\refmark{\men-\pil}.
For example,
  evidence of CP violation in the production of
two Higgs, $H_1$ and $H_2$, in $e^+e^-$ colliders can be obtained
by observing the processes\refmark\men
$$\eqalign{e^+e^-\rightarrow Z^*&\rightarrow ZH_1\, ,\cr
                      &\rightarrow ZH_2\, ,\cr
e^+e^-\rightarrow Z^*&\rightarrow H_1H_2\, .}\eqn\mio$$
Another possibility would be to use polarized photons produced by
back-scattering laser beams at a TeV scale
 $e^+e^-$ collider\refmark{\grzb}.  In this case,
 large Higgs production  asymmetries
in $\gamma\gamma$ collisions
would  provide evidence of CP violation.
\REF\sik{P. Sikivie et al., Nucl. Phys. {\bf B173} (1980) 189.}

In this paper we propose a new way to constrain CP violation in
the Higgs sector. The idea is as follows.
 We know that experimentally   $\rho\equiv
 m^2_W/(m^2_Z\cos^2\theta_W)\simeq 1$. This
can  be understood as a consequence of
an approximate global
$SU(2)$ symmetry of the lagrangian, called in the literature
 a custodial symmetry\refmark\sik.
We will show that when  we have CP
violation in the Higgs sector, such  a custodial $SU(2)$
symmetry cannot be defined
in the Higgs potential.\foot{This was previously pointed out by
S. Weinberg in Ref.~[\wei].}\
Thus, if we insist in having CP violation,
 radiative corrections to
 $\rho$
   will be unavoidable.
{}From the experimental bounds on
  $\Delta\rho\equiv\rho-1$,
  we will be able to
 constrain the CP violation in the Higgs sector.

In section 2, we analyze the scalar contributions to the $\rho$ parameter
in a CP-violating two Higgs doublet model.
  The custodial symmetry  is defined in section 3 where we
prove  our assertion that a custodial-invariant Higgs potential is always
CP conserving.   In section 4
 we use the results of section 2 and 3 in an attempt
to place bounds on CP violation in the Higgs sector. Finally, section 5
is devoted to the conclusions.

\noindent{\bf 2. Contribution to the $\rho$ parameter from a Higgs
 sector with maximal CP violation}

CP violation can occur in the Higgs sector
in models with several scalar multiplets.
Models with only Higgs doublets are particularly
interesting because  at tree level $\rho=1$ is insured in a natural way.
   This is true  whether or not
we impose a custodial symmetry on the scalar potential.
Higher representations, on the other hand,  require a fine-tuning of the
vacuum expectation  values (VEVs) of the neutral scalars
in order that $\rho\simeq 1$.\foot{There are some exotic
 representations, such
   as a 7-plet of hypercharge $Y=\pm 4$,
for which $\rho=1$ at tree level independently of the VEVs of
the scalars.}\
In such
models  the $\rho$ parameter is arbitrary
and radiative corrections are not calculable.
\REF\bra{G.C. Branco, J.-M. Gerard and W. Grimus, Phys. Lett. {\bf B136}
(1984) 383.}

In this paper we will only work within
the two Higgs doublet model (THDM).
Our results, however, can be generalized to most of the multi-scalar
models.
\REF\lee{T.D. Lee, Phys. Rev. {\bf D8} (1973) 1226.}
The most general $SU(2)_L\times U(1)_Y$ gauge invariant
two Higgs doublet potential is
$$\eqalign{V(\Phi_1,\Phi_2)=&\ m^2_{1}\Phi_1^\dagger\Phi_1
+m^2_{2}\Phi_2^\dagger\Phi_2-(m^2_{12}\Phi_1^\dagger\Phi_2+h.c.)\cr
+&\ \lambda_1(\Phi_1^\dagger\Phi_1)^2+
\lambda_2(\Phi_2^\dagger\Phi_2)^2+\lambda_3(\Phi_1^\dagger\Phi_1)(
\Phi_2^\dagger\Phi_2)
+\lambda_4(\Phi_1^\dagger\Phi_2)(\Phi_2^\dagger\Phi_1)\cr
+&\ \coeff{1}{2}\left[\lambda_5(\Phi_1^\dagger\Phi_2)^2+h.c.\right]
+\coeff{1}{2}\left[\Phi_1^\dagger\Phi_2\{  \lambda_6(\Phi_1^
\dagger\Phi_1)+
\lambda_7(\Phi_2^\dagger\Phi_2)\}+h.c.\right]  \, .}\eqn\potential$$
In general
the parameters $m^2_{12}$, $\lambda_5$, $\lambda_6$ and $\lambda_7$
 can be complex and thus give rise to CP violation in the Higgs
  sector.\foot{Strictly speaking, the presence of
   complex scalar self-couplings is
   a necessary but not
sufficient condition  for CP violation\refmark\bra.}\
 Even if all the parameters are real (no
explicit CP violation) it can be shown that there is a region of
parameter space where the VEVs of the neutral scalars are
$$\VEV{\phi^0_1}=v_1\ \ \ ,\ \ \ \VEV{\phi^0_2}=v_2e^{i\xi}\ \ \
 \xi\not= n\pi\ (n\in {\rm\bf Z}), \eqn\veva$$
and therefore CP is  spontaneously broken\refmark\lee.
It is convenient to define a basis $\Phi^\prime_1$, $\Phi^\prime_2$
where only one Higgs doublet gets  a VEV, \ie,
$$\eqalign{\Phi^\prime_1=&\ \cosb\ \Phi_1+\sinb\ e^{-i\xi}\Phi_2=
\left(\matrix{G^+\cr v+\coeff{1}{\sqrt{2}}\left(\hl +iG^0 \right)}
\right)\, ,\cr
\Phi^\prime_2=&-\sinb\ \Phi_1+\cosb\ e^{-i\xi}\Phi_2=
\left(\matrix{H^+\cr \coeff{1}{\sqrt{2}}\left(\hh +i\ha\right)}
\right)\, ,}\eqn\doublets$$
where
$v\equiv \sqrt{v^2_1+v^2_2}$, $\tan\beta=v_2/v_1$,
$G^+$ and  $G^0$ are the Goldstone bosons and  $H^+$ is
the charged Higgs.
 The three
 neutral Higgs boson mass eigenstates $H_{i=1,3}$ are
 mixtures of $\hl$, $\hh$ and $\ha$:
$$\left(\matrix{H_1\cr H_2\cr H_3}\right)
=\bold O\left(\matrix{\hl\cr \hh\cr \ha}\right)\, ,\eqn\ooo$$
where $\bold O$ is an orthogonal matrix.
\REF\tou{D. Toussaint, Phys. Rev. {\bf D18} (1978) 1646.}
\REF\lit{R.S. Lytel, Phys. Rev. {\bf D22} (1980) 505.}
\REF\fre{J.-M. Fr\`ere and J.A.M. Vermaseren, Z. Phys. {\bf C19} (1983)
63.}
\REF\ber{S. Bertolini, Nucl. Phys. {\bf B272} (1986) 77.}
\REF\hol{W. Hollik, Z. Phys. {\bf C32} (1986) 291.}
\REF\fro{C.D. Froggatt, I.G. Knowles and R.G. Moorhouse, Phys. Rev.
 {\bf  D45} (1992) 2471; Nucl. Phys. {\bf B386} (1992) 63.}

Higgs boson contributions to
 the $\rho$ parameter have been extensively analyzed in theories with
 two Higgs doublets without CP violation\refmark{\tou-\hol}\
and  also more recently in the CP-violating case\refmark\fro.
Nevertheless, the connection between the magnitude of
such contributions, the custodial symmetry
 and   CP
violation in the Higgs sector has not been analyzed.
\REF\pes{M.E. Peskin and T. Takeuchi, Phys. Rev. Lett. {\bf 65}
(1990) 964.}
\REF\vel{F. Antonelli, M. Consoli, J. Corbo, Phys. Lett. {\bf B91}
 (1980) 90;
M. Veltman, Phys. Lett. {\bf B91} (1980) 95.}
\REF\lav{L. Lavoura, Int. J. Mod. Phys. {\bf A8} (1993) 375.}
\REF\ken{D.C. Kennedy and P. Langacker, Phys. Rev. Lett. {\bf 65}
(1990) 2967; Phys. Rev. {\bf D44} (1991) 1591.}
\REF\blo{A. Blondel, F.M. Renard and C. Verzegnassi, Phys. Lett.
{\bf B269} (1991) 419.}

The loop contributions to $\Delta\rho$ are given by the parameter $T$
defined by\refmark\pes
$$\alpha T\equiv\coeff{g^2}{m^2_W}
\left[\Pi_{11}(0)-\Pi_{33}(0)\right]\, ,\eqn\deft$$
where $\Pi_{ab}(q^2)$ is the coefficient  of $g^{\mu\nu}$ in the
vacuum polarization tensor and $\alpha=e^2/(4\pi)$.
The Higgs contributions to $T$ are not finite. The gauge boson
  contributions  must also be included to  obtain a  finite
   result. For this reason, it is convenient to set the SM with one
Higgs doublet ($H_{ref}$)
as a reference point and study the deviations from
this  point. In this case, the extra contribution to $T$
in a CP-violating THDM is given by\foot{See also Ref.~[\fro].}
$$\eqalign{ \alpha T&=-\coeff{3g^{\prime 2}}{64\pi^2}
\sum^3_{i=1}\coeff{
\bold O^2_{i1}}{m^2_W-m^2_Z}
L(m^2_{H_i},m^2_{H_{ref}})\cr
&+\coeff{g^2}{64\pi^2 m^2_W}\left[\sum^3_{i=1}\left(1-
\bold O^2_{i1}\right)F(m^2_{H_i},m^2_{H^+})-
\coeff{1}{2}\sum^3_{i,j,k=1\atop i\neq j\neq k}
\bold O^2_{i1}F(m^2_{H_j},m^2_{H_k})\right]\, ,}\eqn\t$$
where
$$L(x,y)=F(x,m^2_Z)-F(x,m^2_W)
+F(y,m^2_W)-F(y,m^2_Z)\, ,\eqn\ddddd$$ and
$$F(x,y)=\coeff{x+y}{2}-\coeff{xy}{x-y}
\ln\coeff{x}{y}\, .\eqn\deff$$
The first term
 reflects the breaking of the
custodial symmetry which results when the $U(1)_Y$ factor is gauged, and
 arises even when the Higgs potential is
custodial-invariant\refmark\vel.
 Such a term grows only logarithmically with
the Higgs masses:
$$\coeff{1}{m^2_W-m^2_Z}L(m^2_{H_i},m^2_{H_{ref}})
\simeq\ln\coeff{m^2_{H_i}}
{m^2_{H_{ref}}}\ \ \ \
\ \ \ {\rm for}\ m_{H_i},\, m_{H_{ref}}\gg m_Z\, .\eqn\log$$
The second term (in brackets) arises, as we will see,
only if the Higgs potential does not have a
 custodial $SU(2)$ symmetry, and
gives a contribution that depends quadratically
on the Higgs masses.
Contrary to what occurs in the CP-conserving  THDM,
this term does not cancel when the $H^+$ is
degenerate with one neutral Higgs boson\refmark{\tou,\lit}.
{}From Eq.~\t\ one can see that this second term vanishes only if
there exists a neutral Higgs boson $H_i$
such that
$$m_{H_i}=m_{H^+}\ \ {\rm and}\ \ \bold O_{i1}=0\, .\eqn\zeros$$
In the case where
 the $H^+$ is degenerate with two neutral
Higgs such  a term also vanishes. It can be easily shown, however,
that when this latter condition holds  Eq.~$\zeros$ also holds.

The relation between CP and the custodial symmetry may be
apparent in Eq.~\zeros  .
It was shown in Ref.~[\men] that in the limit  where
$\bold O_{i1}\rightarrow 0$ (for any $i$) CP is conserved
by  the Higgs-gauge interactions. Consequently, if the
contributions to $T$ are
small  then CP violation is also small. Note that the opposite is not
 true\refmark{\tou-\hol}.
The relation between the custodial and CP symmetry will become
clearer in the next section.
\FIG\figa{Extra contribution to $T$ from a THDM with maximal
CP violation ($\bold O_{i1} = \coeff{\sqrt{3}}{3}$).
The experimental allowed region corresponds to the region
inside the dotted lines.}

In principle there is no reason to expect that
Eq.~\zeros\ holds, so the
contribution to $T$ for large Higgs masses could be significant.
Since $T$ depends on six arbitrary parameters it is
difficult to give a full analysis  of its value.
We will focus on
the contribution to $T$ from a maximally CP-violating Higgs sector.
This occurs
when\refmark\men
$$\bold O_{11}\simeq \bold O_{21}\simeq \bold O_{31}
\simeq\coeff{\sqrt{3}}{3}\, .\eqn\oitres$$
Maximal CP violation  also
requires a large splitting between the
neutral Higgs masses\refmark{\weib,\lav}.
Since it may be more natural for the Higgs boson mass
splitting to be of the order
of their masses, \ie,
$$\Delta m^2_H\sim m^2_H\, ,\eqn\nat$$
we will also consider the case where both Eqs.~\oitres~ and
\nat~ hold.
To give an idea of the orders of magnitude, we have plotted in
 Fig.~\figa\ the parameter
$T$ as a function of the charged Higgs mass for two different
sets
of neutral Higgs masses. In fig.~\figa a we have considered
 a large mass splitting.
 We have chosen
($m_{H_1}, m_{H_2}, m_{H_3}$) =
($60, 500, 1000$) GeV  and ($60, 1000, 1100$) GeV.
In Fig.~\figa b
 we have followed the natural condition of Eq.~\nat\ and chosen
$m_{H_1}=\coeff{1}{2}m_{H_2}=m_{H_3}=\coeff{3}{2}m_{H^+}$ and
$m_{H_1}=\coeff{1}{2}m_{H_2}=\coeff{3}{2}m_{H^+}=2m_{H_3}$.
Following Ref.~[\ken], we have taken
 $m_{H_{ref}}=m_t=m_Z$ as the
 reference point.
{}From Fig.~\figa a one can see that a large mass splitting
results in a large contribution to
$T$. This was expected as we noted before.
When the Higgs mass splitting is  on the order of their masses,
the contribution is only relevant for a large Higgs mass scale
(Fig.~\figa b).
The experimental limits on $T$ (dotted line) have been taken from
 Ref.~[\ken],
$$-0.93<T<0.33\, .\eqn\limit$$
This experimental bound can be used to rule out
a region of parameter space of the Higgs sector.  However,
since the top quark mass is still unknown there is a large
uncertainty in this excluded region
(a large negative contribution to $T$ from a
CP-violating Higgs sector may be canceled by the positive
contribution from the top quark).
Nevertheless, it is likely that the top quark will be discovered
in the near future.
If it were found and its mass measured to an accuracy
of $\Delta m_t\simeq\pm 5$ GeV, it would be possible\refmark\blo,
using the ``Ultimate LEP" configuration,
to improve the experimental uncertainty of $T$ to
$$\Delta T\simeq \pm 0.1\, .\eqn\acc$$
In this case,  if the central value of $T$ remained
close to zero (after subtracting the top quark contribution),
any custodial-breaking term in the Higgs potential would be
tightly constrained.\foot{Actually, due to the first term in Eq.~\t,
a  small experimental value
for $T$ would be required even in models with a custodial-invariant
scalar potential.}\
Of course, we could fine-tune the parameters of the Higgs potential
in order to make the contribution to $T$ small without requiring
an approximate custodial symmetry. We will not consider such an
unnatural possibility.

\noindent{\bf 3. CP and Custodial Symmetry in the Higgs Potential}
\REF\gunc{J.F. Gunion, R. Vega and J. Wudka , Phys. Rev. {\bf D43}
 (1991) 2322.}
\REF\ng{E. Ma and D. Ng, preprint UCRHEP-T103 (1993).}
\REF\hab{H.E. Haber and A. Pomarol, Phys. Lett. {\bf B302} (1993) 435.}

In this section we  study the CP symmetry in the limit where the Higgs
potential is custodial $SU(2)$ invariant.
The custodial symmetry is actually violated
 by the gauge factor $U(1)_Y$  and
infinite radiative corrections\refmark\gunc\  are induced in a
custodial-invariant
Higgs potential
(unless the gauge group is enlarged --see for example Ref.~[\ng]).
Thus, the custodial symmetry can at best  be  only
considered an approximate  symmetry.

We need to enlarge the global symmetry of the Higgs potential
so that
after the ESB  there remains a residual $SU(2)$ symmetry. If
only three Goldstone bosons are  allowed,
the symmetry-breaking pattern is
$$\lr\rightarrow SU(2)_V\, .\eqn\pat$$
In the THDM there are two possible
ways  of defining the global $\lr$ symmetry\refmark\hab:

\noindent
{\bf Case I:} Following the SM case\refmark\sik,
 we define a matrix field
$$M_i=\left(i\tau_2\Phi^*_i\
 \Phi_i\right)\equiv\left(\matrix{\phi^{0*}_i&
\phi^+_i\cr -\phi^-_i&\phi^0_i}\right)\ \ \ \ i=1,2\, ,\eqn\mdu$$
which transforms under the $\lr$ symmetry as
$$M_i\rightarrow L\, M_iR^\dagger\, .\eqn\mlr$$
In order that the symmetry is broken down to $SU(2)_V$, we need
 $$\VEV{M_i}=v_i{\bf 1}\ \ \
 \Rightarrow \ \ \ \VEV{\phi^0_i}=
 \VEV{\phi^{0*}_i}=v_i\in {\rm\bf R}\, .\eqn\breaking$$

\noindent
{\bf Case II:} We
define a unique matrix field
  $M_{21}=\left(i\tau_2\Phi^*_2\ \Phi_1\right)$
which transforms under $\lr$ as
$$M_{21}\rightarrow L\, M_{21}R^\dagger\, .\eqn\mlrb$$
In this case, the symmetry is broken down to $SU(2)_V$ when
$$\VEV{M_{21}}\propto {\bf 1}\ \ \ \Rightarrow\ \ \
\VEV{\phi^0_1}=
\VEV{\phi^{0*}_2}\in {\rm\bf C}\, .\eqn\breakingb$$
It can be shown that, in the absence of fermions,  transformation  \mlrb\
is  a more restrictive  case of transformation \mlr\  .
However, since we will  be considering  the  Higgs-fermion
interactions we need to consider both cases separately.

We begin with case I and define in the usual way
the  transformation of the scalar fields under CP:
$$\eqalign{\Phi_1(\vec x,t)&\rightarrow\Phi^\dagger_1(-\vec x,t)\, ,\cr
\Phi_2(\vec x,t)&\rightarrow\Phi^\dagger_2(-\vec x,t)\, .}\eqn\cpt$$
Note,  that from Eq.~\doublets\ and Eq.~\breaking,  the above
 transformation  is equivalent to defining
 $\hl$ and $\hh$ as CP-even states and $\ha$ as a CP-odd state.
The transformation \cpt\
can be written in the form
$$M_{i}(\vec x,t)\rightarrow \tau_2M_{i}(-\vec x,t)\, \tau_2\ \ \ \
\ \ i=1,2\, .\eqn\cptrans$$
Clearly,  a Higgs potential that is invariant under \mlr\ (with $L=R$)
is also invariant under \cptrans.
Therefore, a Higgs potential that is
invariant under $SU(2)_V$ will automatically
conserve  CP.

For the case II,  we define
the CP transformation as
$$\eqalign{\Phi_1(\vec x,t)&\rightarrow\Phi^\dagger_2(-\vec x,t)\, ,\cr
\Phi_2(\vec x,t)&\rightarrow\Phi^\dagger_1(-\vec x,t)\, .}\eqn\cptb$$
Again this transformation can be written in the form
$M_{21}(\vec x,t)\rightarrow \tau_2M_{21}(-\vec x,t)\, \tau_2$.
It corresponds (using Eq.~\doublets\
and Eq.~\breakingb) to defining
$\hl$ and $\ha$ as CP-even states, and
$\hh$ as CP-odd.

 In the appendix we derive the constraints
 on the most general THDM potential
[Eq.~\potential] imposed by
the  $\lr$ symmetry. We also  show  explicitly that CP is conserved
in a THDM with custodial invariance.

The kinetic term for the Higgs bosons is
invariant under the $\lr$
symmetry only if $g^\prime=0$.  However, for $g^\prime\not= 0$
the transformation
$M\rightarrow \tau_2M\, \tau_2$  ($M=M_i$ or $M_{21}$)
leaves the kinetic term invariant
if the gauge boson  $B^\mu$ associated with the $U(1)_Y$ factor transforms
 as $B^\mu\rightarrow -B^\mu$.
We conclude then that CP
 {\it  is conserved by the gauge-Higgs sector if the
Higgs potential has a custodial $SU(2)$ symmetry.}
\REF\gla{S. Glashow and S. Weinberg, Phys. Rev. {\bf D15} (1977) 1958.}
\REF\weic{S. Weinberg, Phys. Rev. Lett. {\bf 37} (1976) 657.}

Let us now consider the Higgs-fermion interactions.
 The most general Yukawa sector
 induces flavor changing neutral currents (FCNC).
The usual way of suppressing these is by means of a discrete
 symmetry\refmark\gla\ that can be softly broken.
CP conservation in the Yukawa interactions requires
 $\hl$ and $\hh$ to be  CP-even and $\ha$ to be CP-odd.
Such assignments of CP quantum numbers are compatible with the assignments
of case I, but incompatible with  those of case II.
Therefore, case II allows for the  possibility of
having an approximate
custodial $SU(2)$ invariance in the Higgs potential and maximal CP
 violation in Higgs-fermion interactions.
Such a possibility, however, leads to a  tightly
 constrained model (see Eq.~\breakingb\ and appendix):
$$\tan\beta\simeq 1\ ,\ \ m_{\hh}\simeq m_{H^+}\, .\eqn\massequal$$
If the  discrete symmetries\refmark\gla\
 are exact symmetries of the lagrangian,
 the minimal model with
CP violation is the Weinberg three Higgs  doublet model\refmark\weic.
If
an $\lr$ symmetry is imposed on the Higgs potential of this model,
 all the parameters are real.
 Thus, CP is conserved  by the full lagrangian.

\noindent{\bf 4. Constraining CP violation from the experimental
bounds on $T$}
\REF\des{N.G. Deshpande and E. Ma, Phys. Rev. {\bf D18} (1978) 2574.}

Let us consider case I. Custodial $SU(2)$
symmetry requires the CP-odd state $A^0$ to be  a mass eigenstate of
mass equal to the charged Higgs mass
(see appendix). Therefore, the custodial limit is given by
$$\left\{\eqalign{
&\bold O_{31}\rightarrow 0\, ,\cr
&\bold O_{32}\rightarrow 0\, ,\cr
&\Delta m^2_{+3}\equiv m^2_{H^+}-m^2_{H_3}\rightarrow 0\, .}\right.
\eqn\custli$$
The limit
$\bold O_{31},\bold O_{32}\rightarrow 0$ corresponds to $H_3
\rightarrow  A^0$, \ie, CP-conserving limit.
In the limit \custli, the contribution to $T$
 (including only terms with quadratic dependence on the Higgs mass)
can be written as
$$\eqalign{\alpha T&=
\coeff{g^2}{64\pi^2m^2_W}
\sum_{i,j=1\atop i\not= j}^2\bold O^2_{i1}\left[\coeff{1}{2}+
\coeff{m^2_{H_j}}{m^2_{H_j}-m^2_{H^+}}+
\left(\coeff{m^2_{H_j}}{m^2_{H_j}-m^2_{H^+}}\right)^2\ln
\coeff{m^2_{H^+}}{m^2_{H_j}}\right]
\Delta m^2_{+3}\cr
&+\coeff{g^2}{64\pi^2m^2_W}\left[
F(m^2_{H_1},m^2_{H^+})+
F(m^2_{H_2},m^2_{H^+})-
F(m^2_{H_1},m^2_{H_2})\right]\bold O^2_{31}\, .}\eqn\nose$$
Note that when $\Delta m^2_{+3}=0$ (custodial limit of a
 CP-conserving THDM\refmark{\lit,\des}),
  there are still terms that grow quadratically
with the heavy Higgs masses. Explicitly, we have
for $m_{H^+}=m_{H_3}\gg m_{H_1},\, m_{H_2}$,
$$\alpha T\simeq\coeff{g^2m^2_{H^+}}
{64\pi^2m^2_W}\bold O^2_{31}\, .\eqn\tita$$
 The experimental bounds on $T$ can be used to constrain
$\bold O_{31}$ and limit the magnitude of CP violation in
the Higgs-gauge sector. Notice that in fact only two
($\Delta m^2_{+3}$ and $\bold  O_{31}$)
 of the three
custodial-breaking parameters in \custli\ can be constrained
(see  Eq. \nose).
Nevertheless, it seems unnatural for the
parameter $\bold O_{32}$  to be large while
$\Delta m^2_{+3}$ and
 $\bold  O_{31}$
are suppressed.  Thus one expects that in constraining
 $\bold  O_{31}$  one is actually constraining CP violation in the full
Higgs sector.

\FIG\figb{ Extra contribution to $T$ from a THDM where CP violation
is parametrized by $\bold O_{31}$ ($\bold O^{max}_{31}=\sqrt{3}/{3}$).
The experimental allowed region corresponds to the region
inside the dotted lines.}

To get an idea of the magnitude of such bounds, we parameterize
the custodial limit \custli\
in terms of only one parameter, $\bold O_{31}$.
 In a first
case (Fig.~\figb a),
 we consider:
$$\eqalign{\bold O_{32}&=\bold O_{31}\, ,\cr
m_{H_3}&=\left(1+\coeff{1}{2}
\coeff{\bold O_{31}}{\bold O^{max}_{31}}\right)
m_{H^+}\, ,}\eqn\custlib$$
where $\bold O^{max}_{31}\equiv\sqrt{3}/{3}$ is the value of
 $\bold O_{31}$ for maximal CP violation\refmark\men. In a second case
(Fig.~\figb b) we again consider, $\bold O_{32}=\bold O_{31}$ and in
 addition
 saturate the third limit
in \custli\ ($m_{H^+}=m_{H_3}$). In both cases we also assume
 $\bold O_{21}\simeq\bold O_{11}$ and
$m_{H_1}=\coeff{1}{2}m_{H_2}=\coeff{3}{2}m_{H^+}$.

When we study Fig.~\figb\   we see that unfortunately the actual
constraints are very sensitive to the different parameters of the model.
The results  are also very sensitive  to how
the custodial limit is taken.
If we fix the ratio
$\Delta m^2_{+3}/\bold  O_{31}$  in
the custodial limit \custli,
 the dependence of $T$ on
$\bold O_{31}$ is linear and then the bounds are tight (Fig.~\figb a).
However, if
$\Delta m^2_{+3}/\bold  O_{31}$ goes to zero in the
custodial limit \custli, then
 $T$ depends quadratically on
$\bold O_{31}$  and the bounds are  less restrictive (Fig.~\figb b).

\noindent{\bf 5. Conclusions}

We have shown that in a THDM with maximal CP violation in the Higgs
sector, there are large contributions to  $\Delta\rho$ coming
from the scalar sector. We have seen that this is due to the fact that
any term in the Higgs potential that breaks CP also
breaks the custodial $SU(2)$ symmetry and, therefore,
contributes  to $\Delta\rho$  at one-loop.
This is true in the Higgs-gauge sector
and can be generalized to models with a more extended
 Higgs sector.

We have used the experimental bounds on the parameter $T$ to
constrain  CP violation in the THDM Higgs sector.
These constraints (Fig.~\figb) are found  to be very
dependent on the parameters of the model.
At present,  they  are the only bounds on
CP nonconservation in the Higgs sector.
When the limits on $T$ improve\refmark\blo\
 ($\Delta T=\pm 0.1$), the region
of parameter space in the Higgs potential that allows for CP  violation
will be further constrained.

It is important to remark that
 when the Higgs masses are large the
 constraints on CP
nonconservation are stronger.  Note that this is opposite to the case
where CP nonconservation is constrained using the electric dipole
 moment.

When Higgs-fermion interactions are  considered,
we have shown that it is possible to find a definition of
the custodial symmetry which allows for CP violation.
Such a possibility
 leads to other interesting constraints  (Eq.~\massequal).

\vskip .5cm
\centerline{\bf Acknowledgements}
We gratefully acknowledge conversations with
 Jack Gunion, Howard Haber and
 Michael Peskin.  We would like to thank Cliff Burgess and Dave Robertson for a
critical reading of the manuscript.   The work of A.P.
 was supported  by a fellowship of the MEC (Spain).

\appendix

In this appendix we derive the restrictions that
the $\lr$ symmetry places on
the most general THDM potential of Eq.~\potential.

For  case I, the most general  $\lr$ symmetric potential is
given by
$$\eqalign{V(M_1,\, M_2)&=\coeff{1}{2}m^2_1{\rm Tr}\{M_1^\dagger M_1\}
+\coeff{1}{2} m^2_2{\rm Tr}\{M_2^\dagger M_2\}-
m^2_{12}{\rm Tr}\{M_1^\dagger M_2\}\cr
&+\coeff{1}{4}\lambda_1{\rm Tr}^2\{M_1^\dagger M_1\}+
\coeff{1}{4}\lambda_2{\rm Tr}^2\{M_2^\dagger M_2\}\cr
&+\coeff{1}{4}
\lambda_3{\rm Tr}\{M_1^\dagger M_1\}{\rm Tr}\{M_2^\dagger M_2\}+
\coeff{1}{2}\lambda{\rm Tr}^2\{M_1^\dagger M_2\}\cr
&+\coeff{1}{4}
{\rm Tr}\{M_1^\dagger M_2\}\left[\lambda_6{\rm Tr}\{M_1^\dagger M_1\}+
\lambda_7{\rm Tr}\{M_2^\dagger M_2\}\right]\, .}\eqn\potentialb$$
Using
 Eq.~\mdu, it is easy to see that the above potential
  corresponds to  Eq.~\potential\ with
 $$m^2_{12}\, ,\,\lambda_6\, ,\,\lambda_7\in {\rm\bf R}\ \ ,\ \
 \lambda=\lambda_4=\lambda_5\in {\rm\bf R}\, .\eqn\cona$$
Thus, all parameters of the Higgs sector are real (see
 Eq.~\breaking\ and Eq.~\cona), and CP is conserved.
Explicitly, we find that  $A^0$ is a mass eigenstate with
  $m_{\ha}=m_{H^+}$.
As we expected,  Eq.~\zeros\ holds and, then the
  contribution to $T$ that grows quadratically with the Higgs masses
   vanishes.

   For case II, the most general potential that is invariant
   under $\lr$ is given by
$$\eqalign{
V(M_{21})&=m^2{\rm Tr}\{M_{21}^\dagger M_{21}\}
-(m_{12}^2\det M^\dagger_{21}+
h.c.)\cr
&+\lambda{\rm Tr}^2\{M_{21}^\dagger M_{21}\}+
\lambda_4\det\{M_{21}^\dagger M_{21}\}+
\coeff{1}{2}\left[\lambda_5\det\{ M_{21}^\dagger\}^2+h.c.\right]\cr
&+\coeff{1}{2}\left[\lambda^\prime\det M^\dagger_{21}\,
{\rm Tr}\{M_{21}^\dagger M_{21}\}+h.c.\right]\, .}
\eqn\potentialc$$
This potential corresponds to  Eq.~\potential\ with
$$m^2=
m^2_{1}=m^2_{2}\ \ ,\ \ \lambda=
\lambda_1=\lambda_2=\coeff{1}{2}\lambda_3\ \
,\ \ \lambda^\prime=
\lambda_6=\lambda_7\ .\eqn\limb$$
It seems that in this case we can have explicit and spontaneous CP
violation, because the scalar self-couplings and the
VEVs  can be
complex. Nevertheless, using Eqs.~\breakingb\ and \limb\ in
 Eq.~\potential,   one can
see that  $H^0$ is a mass eigenstate (with $m_{H^0}=m_{H^+}$).
Therefore, considering only the Higgs-gauge sector, we can
 define\refmark\men\
$H^0$ to be  CP-odd and the other neutral Higgs  to be CP-even.
\vskip .5cm
\endpage
\refout
\endpage
\figout
\end